\documentclass{article}
\usepackage{graphicx}
\usepackage{amsmath}
\usepackage{color}
\usepackage[a4paper]{geometry}
\usepackage{parskip}

\title{Core-biased random walks in complex networks.}
\author{Ra\'ul J. Mondrag\'on\\School of Electronic Engineering and Computer Science\\Queen Mary University of London\\ Mile End Road E1 4NS UK}
\date{20$^{\rm th}$ September 2017}

\def\degree{{k}}
\def\linksRC{k^+}

\def\degree{k}

\def\walk{W}
\def\jumpProb{P_{i\rightarrow j}}
\def\firstEigenvalue{\Lambda}
\def\firstEigenvector{\overline{v}}

\begin{document}
\maketitle

\section*{Abstract}
A simple strategy to explore a network is to use a random-walk where the walker jumps from one node to an adjacent node at random.  It is known that  biasing the random jump, the walker can explore every walk of the same length with equal probability, this is known as a Maximal Entropy Random Walk (MERW).  To construct a MERW requires the knowledge of the largest eigenvalue $\firstEigenvalue$ and corresponding eigenvector $\firstEigenvector$ of the adjacency matrix ${\bf A} = \{a_{ij}\}$, this requires global knowledge of the network.  When this global information is not available, it is possible to construct a biased random walk which approximates the MERW using only the degree of the nodes, a local property. Here we show that it is also possible to construct a good approximation to a MERW by biasing the random walk via the properties of the network's core, which is a mesoscale property of the network. We present some examples showing that the core-biased random walk outperforms the degree-biased random walks.

\section{Introduction}
A simple method to explore a complex network is via a random walk. Properties of the random walk, like the probability that the walker is in a particular node as  time tends to infinity or, the time that it would take the walker to visit at least once all the nodes of the network, are dependent on the network's connectivity.  This relationship between the structural properties of the network and the random walk have been used to explore properties of complex networks like community detection~\cite{rosvall2008maps}.

Here we are interested in biased random walks, and in particular a random walk where the walkers explore every walk with equal probability, known as Maximal Entropy-rate Random Walks (MERW)~\cite{burda2009localization}. This biased random walks are used to study well known problems in complex networks like link prediction~\cite{li2011link} but also in practical applications, for example in the discovery of salient objects in an image~\cite{yu2014maximal}. To construct a MERW requires the largest eigenvalue and corresponding eigenvector of the adjacency matrix~\cite{burda2009localization} which are global properties of the network. 

A good approximation to the MERW using only local properties is the degree-biased random walk~\cite{gomez2008entropy,fronczak2009biased}. In this case the walker jumps from one node to a neighbouring node with a preference based on the degree of the destination node. This degree-biased random walks are also used to understand properties of networks, for example how to efficiently disperse the information contained in a node on a network~\cite{gomez2008entropy}, 
in the link prediction problem~\cite{liu2010link} and more recently the degree-biased random walk has been extended to multiplex networks~\cite{battiston2016efficient}.

In here, as an approximation to the MERW, we present a biased random walk where the walker prefers to jump to nodes that are connected to the core of the network. We would justify the proposed biased random walk via an spectral bound of the adjacency matrix and show that this core-biased random walk can outperform the degree-biased random walk.

The connectivity of a finite, undirected and connected network can be described by the symmetric adjacency matrix ${\bf A}$, where $a_{ij} = 1$ if nodes $i$ and $j$ share a link and zero otherwise. In these networks, a random walker would jump from node $i$ to  a neighbouring node $j$ with a probability $\jumpProb$. The probability that the walker is in node $j$ at time $t+1$ is $p_j(t+1) = \sum_i a_{ij}\jumpProb p_i(t)$ or in matrix notation $\overline{p}(t+1)=\boldsymbol{\pi} \overline{p}(t)$. If the matrix $\boldsymbol{\pi}$ is primitive then the probability of finding the walker in node $i$ as the times tend to infinity is given by the stationary distribution $\overline{p}^*=\{p^*_i\}$. 
In a network, the jump probability $\jumpProb$ can be expressed as
\begin{equation}
\jumpProb = \frac{a_{ij}f_j}{\sum_j a_{ij}f_j}
\label{eq:defJump}
\end{equation} 
where $f_j$ is a function of one or several topological properties of the network, in this case the stationary distribution is~\cite{gomez2008entropy}
\begin{equation}
p^*_i=\frac{f_i\sum_j a_{ij}f_j}{\sum_n f_n\sum_j a_{nj}f_j}.
\label{eq:statProb}
\end{equation}

The measure which tell us the minimum amount of information needed to describe  the stochastic walk is the entropy rate $h=\lim_{t\rightarrow\infty}S_t/t$, where $S_t$ is the Shannon entropy of all the walks of length $t$. This entropy is related to the properties of the random-walk via~\cite{burda2009localization}
\begin{equation}
h=-\sum_{i,j}
p^*_i\jumpProb  \ln (\jumpProb).
\label{eq:entrop}
\end{equation}
The maximal entropy rate $h_{\rm max}$ corresponds to random walks where all the walks of the same length have equal probability.
The value of $h_{\rm max}$ is related to the spectral properties of the network. 
If $\firstEigenvalue$  is the largest eigenvalue and $\firstEigenvector$ its corresponding eigenvector of the adjacency matrix then the maximal entropy satisfies $h_{\rm max}= \ln (\Lambda)$~\cite{burda2009localization}.
 The Maximal Entropy Random Walk (MERW)  is obtained when the transition probability from node $i$ to node $j$ is 
$\jumpProb=a_{ij}v_j/(\sum_j a_{ij}v_j)$~\cite{burda2009localization}, where $v_i$ is the $i$--th entry of the eigenvector $\firstEigenvector$. 

The implementation of the MERW requires global knowledge of the network connectivity as we need to evaluate the largest eigenvalue-eigenvector pair. If this global knowledge is unavailable then it is possible to approximate the maximal entropy random walk via the node's degree $k_i$~\cite{gomez2008entropy,fronczak2009biased} which is a local network property. The case $f_j=\degree_j^\alpha$, where $\degree_j$ is the degree of node $j$ and $\alpha$ a free parameter has been extensively studied~\cite{bonaventura2014characteristic,sinatra2011maximal,battiston2016efficient}. In this case, from  Eq.~(\ref{eq:defJump}), the jump probability is
\begin{equation}
\jumpProb = \frac{a_{ij}\degree_j^\alpha}{\sum_j a_{ij}\degree_j^\alpha}.
\label{eq:vito1}
\end{equation}
The case $\alpha=0$ corresponds to the unbiased random walk where $\jumpProb = a_{ij}/(\sum_j a_{ij})$ and if $\alpha=1$ the jump is proportional to the node's degree.

It is possible to include more topological information in the biased random jump to improve the approximation to the MERW. For example adding the information about the neighbours average degree $\degree^{(1)}_j$ of node $j$, in this case the random jump probability is~\cite{sinatra2011maximal}
\begin{equation}
\jumpProb= \frac{a_{ij}\degree_j \degree_j^{(1)}}{\sum_j a_{ij}\degree_j\degree_j^{(1)}}.
\label{eq:vito2}
\end{equation}
\section{Core-biased random walk in a network.}
If the largest eigenvalue-eigenvector pair is not known, the MERW results suggests that a good approximation to the largest eigenvector $\firstEigenvector$ could be used to construct an efficient biased random walk. 
In here, we are going to use a lower bound for $\firstEigenvalue$ based on the network's core to construct a biased random walk.

A lower bound for $\firstEigenvalue$ is \cite{van2010graph} $\firstEigenvalue \ge (\walk_{2n}/\walk_0)^{1/(2n)}\ge (\walk_n/\walk_0)^{1/n}$, $n=1,\ldots$ where $\walk_n = \overline{u}^T{\bf A}^n \overline{u}$ is the total number of walks of length $n$, ${\bf A}$ is the adjacency matrix and $\overline{u}$ is a vector with all its entries equal to one. A lower bound for the number of walks is $\walk_n\le \sum_i^N \degree_i^n$~\cite{fiol2009number} where the equality is true only if $n \le 2$.

In a network where the nodes are ranked in decreasing order of their degrees, the connectivity of the network can be described with the degree sequence $\{\degree_r\}$ and the sequence $\{ \linksRC_r \}$ where $\linksRC_r$ is the number of links that node $r$ shares with nodes of higher rank. 
A bound to the largest eigenvalue in terms of the $\{\linksRC_r\}$ sequence is $\firstEigenvalue\ge 2\langle \linksRC \rangle_{{r}}$, where  $\langle\linksRC\rangle_r=(1/r)\sum_i^r\linksRC_i$ is the average number of links shared by the top ranked $r$ nodes~\cite{mondragon2016network}. The core is defined by the rank  $r$ where $\langle \linksRC \rangle_{{r}}$ is maximal.  Any node with rank greater or equal to $r$ belongs to the core. If $r=N$ then we recover the well known bound $\firstEigenvalue \ge \walk_1/\walk_0=2(\sum_i^N\linksRC_i)/N=L/N$ where $\walk_0=N$ is the total number of nodes and $\walk_1=L$ is the total number of links.

Consider the adjacency matrix ${\bf A}$ of the network where the nodes are ranked in decreasing order of their degrees and $\overline{u}(r)$ a vector where its first $r$ entries are set to one and the rest to zero.  The vector $\overline{z}(r)={\bf A}\overline{u}(r)$ has entries $z_i(r)=K^+_i(r)$, where $K^+_i(r)$ is the number of links that node $i$ shares with the top $r$ ranked nodes. Also notice that $\langle \linksRC \rangle_{{r}}=(\overline{u}^T(r)\,{\bf A}\,\overline{u}(r))/r$.  As $2\langle \linksRC \rangle_{{r}}$  is a lower approximation to $\firstEigenvalue$,  then $\overline{z}(r)$ gives an approximation to the eigenvector $\firstEigenvector$.

This bound based on $\{\linksRC_i\}$ suggest a biased random walk based on the core. If the top $r$ ranked nodes, that is nodes $\{1,\ldots,r\}$, are the core of the network, then a core-biased random jump is $P_{i\rightarrow j} =  a_{ij}K^+_j(r)/\sum_j (a_{ij}K^+_j(r))$. But it is possible that $K^+_j(r)=0$ if node $j$ has no links with the network's core and then the random-walk will be ill-defined. To avoid this situation we propose 
\begin{equation}
P_{i\rightarrow j} =  \frac{a_{ij}(K^+_j(r)+1)}{\sum_j^N a_{ij}(K^+_j(r)+1)}.
\label{eq:biasedCore}
\end{equation}
Practically to implement this method, the random-walker needs to know the number of neighbours that node $j$ has with the core. 
The number of neighbours of node $j$  is a local property but, to measure how many of the neighbours are in the core  requires the rank of the neighbouring nodes and the value of  $r$, which are global properties.  
However, at the end of this article, we would propose how to extend the core-biased random walk even when we don't know which nodes form the core of the network.


\section{Examples from real networks}
\begin{figure}[t]
\begin{center}
\includegraphics[width=\textwidth]{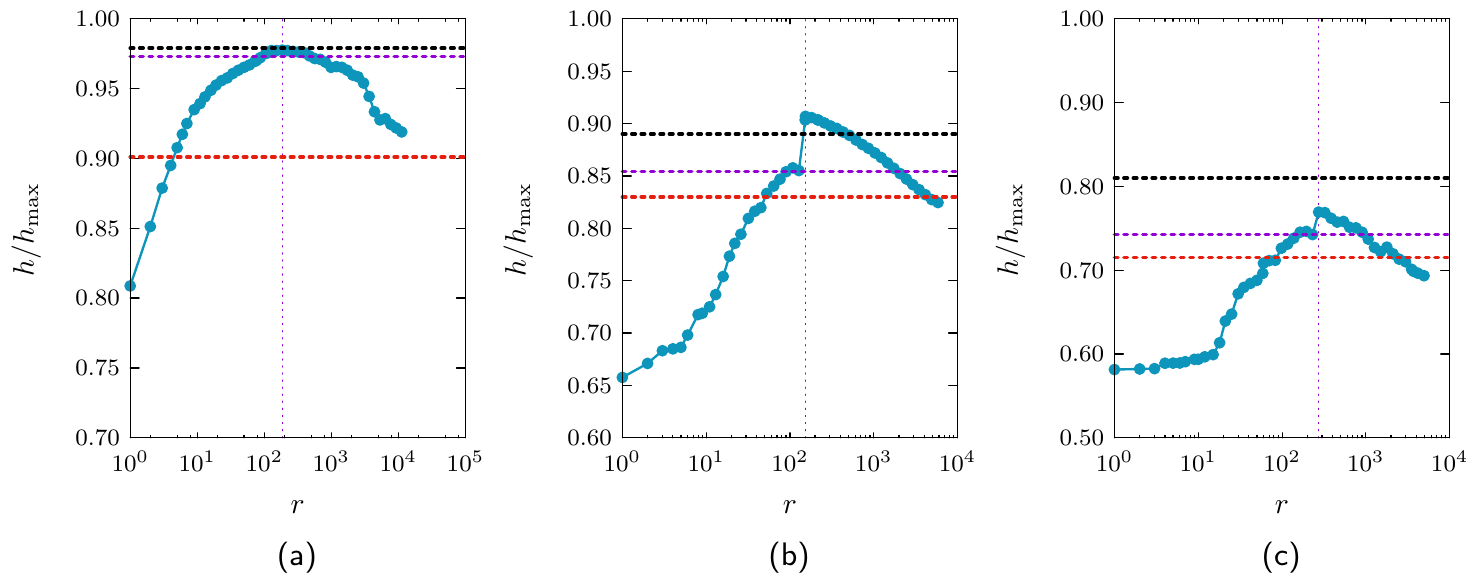}
\end{center}
\caption{\label{fig:networks} Ratio $h/h_{\rm max}$ for (a) the AS-Internet, (b) HepTh and (c) Power networks. The vertical lines show the value of $r$ when $h$ is maximal. The horizontal lines are different maximal values of $h$ obtained from Eq.~(\ref{eq:vito1}) (bottom and middle)  and Eq.~(\ref{eq:vito2}) (top line). The middle horizontal line was obtained when $\alpha=0.64$, $\alpha=1.39$ and $\alpha=1.60$ which maximises the value of $h$ for the AS-Internet, HepTh and Power networks respectively.}
\end{figure}

We consider three real networks, the AS-Internet, the network of co-authors for Physicist researching in High Energy Physics (Hep-Th) and an electrical distribution network (Power network). Figure~\ref{fig:networks} shows $h/h_{\rm max}$ against the size of the core $r$ for these networks. The value of $h$ was evaluated from the core biased random jump $\jumpProb$ given by Eq.~(\ref{eq:biasedCore}), 
then evaluating the stationary distribution $\{p^*_i\}$ (Eq.~\ref{eq:statProb}) and from this distribution the rate entropy $h$ (Eq.~(\ref{eq:entrop})). The maximal entropy $h_{max}=\ln \firstEigenvalue$ was evaluated from the diagonalisation of the adjacency matrix ${\bf A}$. In the figure, for comparison purposes, the bottom horizontal line shows the ratio $h/h_{\rm max}$ for the degree-biased random walk ($\alpha=1$), the middle horizontal line for the degree-biased random walk where the value of $\alpha$ maximises $h$ and the top horizontal line the case when the average neighbours degree is included (Eq.~(\ref{eq:vito2})).

In the figure the joined blue dots  show the ratio $h/h_{\rm max}$ against the size of the core for the core-biased random jump given by Eq.~(\ref{eq:biasedCore}). For the three networks there is a large interval in the rank $r$ where the core-biased random walk performs better than the degree-biased random walk (bottom and middle horizontal lines), and in the Internet and HepTh network there are values of $r$ where the  core-biased random walk is better than the method considering the average neighbours degree (top line). For the Power network the approximation to $h_{max}$ is not as good as for the AS-Internet or the HepTh. The reason is that the vector $\overline{z}(r)$ with entries $z_i(r)=K^+_i(r)$ gives a poor approximation to the eigenvector $\firstEigenvector$~\cite{mondragon2016network}. In this case the entries $v_j$ which correspond to nodes with middle to low degree have relatively large values. 

The vertical lines show the rank $r$ where $h$ is maximal. If we define the core of the network at this value then the core size is relatively small. The percentage of nodes relatively to the size of the network that form the core is 1.6\% for the AS-Internet, 2.6\% for the HepTh and 5.5\% for the Power network.

\section{Concluding remarks}
The justification  behind the proposed core-biased random walk is the bound $\firstEigenvalue\ge 2\langle \linksRC \rangle_r$ which relates the connectivity of the high degree nodes with the spectral properties of the network. In the core-biased jump the walker prefers to jump to the nodes of high degree that share links with other nodes of high degree.

The sequence $\{\linksRC_i\}$, describing the connectivity between node $i$ with nodes of higher rank, is relevant when describing undirected networks as it has been used to evaluate, in closed form the maximal entropy ensemble~\cite{mondragon2014network}, the definition of the rich-core~\cite{ma2015rich}, the spectral-core~\cite{mondragon2016network} and now biased random walks.
 
Finally, the core-biased random walk gives a good approximation to the maximal entropy random walk, however, it requires the ranking of the nodes by their degree.
From the real network examples presented here, it is clear that there is a large interval in the rank $r$ where the core-biased random walk performs better than the degree-biased random walk. This suggest that it is possible to create an efficient biased random walk even when the overall ranking of the nodes it is not known. If we assume that the core are the nodes with degree greater than $k^*$ and $k^*<k_{\rm max}$ where $k_{\rm max}$ is the maximal degree  of the network, then we can redefine $K^+_j(k^*)$ as the number of links that node $j$ has with nodes of degree higher or equal to $k^*$ and $\jumpProb =  (a_{ij}(K^+_j(k^*)+1))/(\sum_j a_{ij}(K^+_j(k^*)+1))$ could give an efficient biased random walk without having to evaluate the ranking of the nodes.


{\small
\bibliographystyle{unsrt}

}
\end{document}